\documentstyle[12pt,epsfig]{article}
\def\nue{\nu_{\rm e}}
\def\num{\nu_\mu}
\def\nut{\nu_\tau}
\def\nus{\nu_{\rm s}}
\def\nubarnu{\stackrel{\makebox[0pt][l]
              {$\hskip-2pt\scriptscriptstyle(-)$}}{\nu}} 
\def\osc{\leftrightarrow}
\def\effe{F_{\rm e}}
\def\effm{F_\mu}
\def\efft{F_\tau}
\def\effa{F_a}
\def\enne{N_{\rm e}}

\def\ennp{N_{\pi^0}}
\def\nc{{NC}}
\def\cce{{CC{\rm e}}}
\def\ccm{{CC\mu}}
\def\rrmue{{\cal R}_{\mu/{\rm e}}}
\def\rrnue{{\cal R}_{\nu/{\rm e}}}
\def\rmue{{R}_{\mu/{\rm e}}}
\def\rpie{R_{\pi^0/{\rm e}}}
\def\rpimm{R_{\pi^0/MM}}
\def\rpism{R_{\pi^0/SM}}
\def\rpimr{R_{\pi^0/{\rm m.r.}}}
\def\rmmsm{R_{MM/SM}}
\begin{document}
\title{Neutral-to-charged current\\ events ratio
in atmospheric neutrinos\\ and neutrino oscillations} 
\author{Francesco Vissani$^{a,b}$ and Alexei Yu.\ Smirnov$^{a,c}$\\ \\
{\it $^a$International Centre for Theoretical Physics, ICTP,}\\
{\it Strada Costiera 11, 34013 Trieste, Italy.} \\
{\it $^b$Deutsches Elektronen-Synchrotron, DESY,}\\
{\it Notkestra\ss{}e 85, 22603 Hamburg, Germany.}\\
{\it $^c$Institute for Nuclear Research,}\\
{\it Russian Academy of Sciences, 107370 Moskow, Russia.}}
\date{}
\maketitle
\begin{abstract}
The atmospheric neutrinos produce 
isolated neutral pions  ($\pi^0$-{\em events}) 
mainly in the neutral current 
interactions. 
We propose to study the  ratios involving
$\pi^0$-{\em events}  and the events induced mainly by 
the charged currents. This 
minimizes uncertainties related to  the original  
neutrino fluxes, and in certain cases,  also to 
the cross-sections. 
Experimental study of these ratios 
will allow one to check the oscillation solution of the 
atmospheric neutrino problem 
and to identify the channel of
oscillations.  Illustrative analysis 
of existing data is presented.
\end{abstract}
\newpage

\section{Introduction}

Recent results from the Super-Kamiokande \cite{SK}
and the SOUDAN 2 experiments \cite{SOUDAN}
have confirmed the   existence of the atmospheric
neutrino problem \cite{prob1,prob2}. 
At the moment, it seems 
that neutrino oscillations are the only  
satisfactory solution of the problem  \cite{atm1,atm2,atm3}.  
Various possibilities of description of the data still 
exist depending on channel 
of oscillations.  
The $\num \osc \nut$ oscillations    
give   the most favorable solution \cite{atm1,atm2,atm3}. 
The  $\num \osc \nue$ channel is  restricted by the 
reactor data \cite{reac1,reac2, reac3}, 
especially by recent results from CHOOZ experiment 
\cite{reac3} and it seems strongly
disfavored by 
recent Super-Kamiokande results on the zenith  angle dependence 
of the e-like and $\mu$-like events. 
At low energies the oscillations to sterile neutrinos,   
$\num\osc\nus$, \cite{steril,steril2} 
lead to  results which are  
similar to  those in the 
$\num\osc\nut$ channel. 
The  $\num\osc\nus$ channel 
is of interest since it gives 
an additional freedom in  reconciliation of  
different  neutrino anomalies. 
Moreover, it allows one to keep 
usual hierarchical mass structure 
and small mixings for active 
neutrinos, while solving  the atmospheric 
neutrino problem  \cite{smi,liu}. 
Combined effect of different channels 
of oscillations is also possible \cite{combine}.

Still a number of  questions exists concerning the  
interpretation of the atmospheric neutrino data.  
They  are related 
to consistency of the experimental results 
and to possible   systematic errors. Therefore, 
the  crucial task is to find new independent criteria 
for cross-checking  the oscillation hypothesis, as well as for 
discrimination of different channels of oscillations. 
Recently several new attempts  
have been made  along with this line  \cite{pakv}. 

Up to now  studies of  the atmospheric neutrinos were   
restricted, mainly,  by events induced by   
the charged current ($CC$) 
interactions  (e-like and $\mu$-like events) 
which compose  the bulk of the data.  
However, new high statistics  experiments:   
the Super-Kamiokande, and later ICARUS,   will allow one to 
get important information from samples of more rare events.  

In this paper we will consider the  possibility to 
use events induced by the {\it neutral currents}. 
The ratio of  
the neutral and charged currents events  
($NC/CC$) can give an important information  
on the neutrino flavor oscillations. 
In the next Section we 
will consider properties of the 
neutrino fluxes relevant for the $NC$ and $CC$ reactions.  
In experiment one observes events of different type 
which do not correspond to  certain  
$NC$ or $CC$ reactions. In this connection 
we identify  (Sect.\ 3)  
the observables which represent closely 
the ratio $NC/CC$ and study their sensitivity to different 
channels of oscillations.  In Sect.\ 4 we  
confront predictions with the available 
Siper-Kamiokande data.

\section{Flavour content of atmospheric neutrinos}

In what follows we will consider  the (relative) fluxes,  
$\effe$, 
$\effm$, and $\efft$ of the $\nue$, $\num$, and $\nut$ 
neutrinos (and corresponding antineutrinos) with energies around 1 GeV 
that induce the ``fully contained'' events   
in the underground detectors. 

The neutral current effects are determined by the total 
flux of the active neutrinos in the detector: 
$\effa \equiv \effe + \effm +\efft .$ 
For the ratio of this flux and the flux 
in absence of oscillations, 
$\effa^0 \equiv \effe^0 + 
\effm^0$ (the predicted $\nut$-flux 
is negligibly small),  
we can write    
\begin{equation}
\frac{\effa}{\effa^0}
 =  \left\{ 
\begin{array}{cr}
1 & \ \ \ \ \ \ \ \ \ \ \ \num \osc \nue \ \ 
           {\rm and} \ \ \num \osc \nut\\[1ex]
\displaystyle{\frac{1+r\, P}{1 + r}} & \num \osc \nus 
\end{array}~,
\right.
\label{ncrate}
\end{equation}
where $r \equiv \effm^0/\effe^0 \approx 2.1$ (in the energy
range  $E_\nu \sim 1$ GeV), and 
$P \equiv P(\num  \osc \num) \leq 1$ 
is the  $\num$ survival probability averaged over the appropriate 
energy range.  
The ratio ${\effa}/{\effa^0}$ 
equals one in the  no-oscillation case as well as  in the case of
flavor oscillations. 
It is smaller than one,  if active neutrinos 
oscillate into sterile neutrinos, 
$\nue \osc \nus .$

The rates of the charged current events with appearance  
of the electrons or muons determine     
$\nue$ or  $\num$ fluxes  separately\footnote{The 
$\tau$ lepton production by the oscillation-induced $\nut$ 
is strongly  suppressed because of the high 
energy threshold $E_{\tau}^{th}$=3.4  GeV.}.
The ratios of fluxes with and without oscillations equal: 
\begin{equation}
\frac{\effe}{\effe^0}
 =  \left\{ 
\begin{array}{cr}
1           &\ \ \ \ \ \ \ \num \osc \nut \ \  
             {\rm and} \ \  \num \osc \nus\\[1ex]
r - (r- 1)\ P & \num \osc \nue 
\end{array}~,
\right.
\label{ncrate1}
\end{equation}
and 
\begin{equation}
\frac{\effm}{\effm^0}
 =  \left\{ 
\begin{array}{cr}
P  & \ \ \ \ \ \ \ \ \  \num \osc \nut \ \ 
    {\rm and} \ \ \num \osc \nus\\[1ex]
\displaystyle{ r^{-1} - (r^{-1} - 1)\ P} & \num \osc \nue 
\end{array}~.
\right.
\label{ncrate2}
\end{equation}

To avoid uncertainties related to the absolute 
normalization of theoretical fluxes 
let us introduce  the double ratio
$\rrnue$:  
\begin{equation}
\rrnue\equiv \frac{(\effe + \effm + \efft)/ 
\, (\effe^0 + \effm^0)}{\effe\,/\,\effe^0}  
\label{rnue0}
\end{equation}
that compares the flux of active neutrinos
with the flux of electron neutrinos at the detector 
normalized on  the same ratio of fluxes 
without oscillations.  
Obviously,   $\rrnue = 1$ in the absence of oscillations. 
We can rewrite the double ratio as  
\begin{equation}
\rrnue   = \frac{1+r\cdot \rrmue + \efft/\effe}{1+r}, 
\label{rnue1}
\end{equation} 
where
$$
\rrmue \equiv (\effm)/\effe)\,/\,(\effm^0/\effe^0)  
$$ 
is  the double ratio for the muon and electron neutrino fluxes. 
Experiment gives the double ratio of the $\mu$-like 
and the  e-like events 
$$
\rmue=(\mu/{\rm e})_{data}\,/\,(\mu/{\rm e})_{MC} = 0.61 \pm 0.05 
$$   
instead of the expected value  1 which, as is known,  composes the 
atmospheric neutrino problem. From this, 
up to small corrections due to
misidentification of events, we can take 
$\rrmue \approx \rmue \approx 0.6 .$  

The double ratio $\rrnue$ (\ref{rnue0}) allows one to trace 
the effect of oscillations on the $NC/CC$ ratios. 
Notice that once $\rrmue$ is fixed by  experiment, the ratio 
$\rrnue$ depends on $\efft/\effe$ only. 
In particular, the minimal value of $\rrnue$ is 
attained for $\efft = 0 .$ 
The ratio $\rrnue$ increases with
the flux of tau neutrinos in the detector (if $\effe$ does not change). 
Using the definition of the $\rrnue$
(\ref{rnue0}) and assuming 
$\effa \le \effa^0,$
we get the 
inequality $\rrnue \le  \effe^0/\effe .$   
In turn, the ratio  $ \effe^0/\effe$  
has at least  20\% theoretical 
uncertainties, and its values smaller than unity  
(but compatible with unity) are experimentally 
preferred. This leads to 
the inequality 
$\rrnue {\ \raisebox{-.4ex}{\rlap{$\sim$}} 
\raisebox{.4ex}{$<$}\ } 1.$

Let us consider the influence of oscillations 
which give a solution of the atmospheric neutrino problem 
(that is, reproduce $\rrmue = 0.6$) 
on the double ratio $\rrnue .$ 
The  $\num \osc \nut$ oscillations  
imply  that $\effe = \effe^0,$ and the total flux of 
active neutrinos is unaffected;  
therefore we have $\rrnue=1.$

In the case of the $\num \osc \nue$ and $\num \osc \nus$
oscillation solutions no $\nut$-flux is produced, and according to 
(\ref{rnue1}) we get 
\begin{equation}
\rrnue   = \frac{1+r\cdot \rrmue}{1+r} = 0.73. 
\label{rnue2}
\end{equation}
That is,  $\rrnue$ reaches its minimal value. 
Let us underline that $\num \osc \nue$ and $\num \osc \nus$ 
oscillations as solutions of the atmospheric problem 
lead to the same  value of ratio  $\rrnue$, although they 
modify  neutral current effects differently.

In the general case, when 
both $\num \osc \nut$ and $\num \osc \nus$ 
 channels contribute to the deficit of the $\num$-flux simultaneously, 
the ratio $\rrnue$ can take any value in the interval 
\begin{equation}
0.73 \le \rrnue \le 1 . 
\label{range}
\end{equation}
Here the right (left) value would signal
the presence (absence) of an induced flux of 
tau neutrinos
(for sufficiently small $\effe/\efft,$
the ratio in formula (\ref{rnue1})
can be even bigger than one, 
although this possibility is disfavored 
experimentally, as discussed above.)  
{}From (\ref{range}) we conclude that better than 
10\% accuracy in measurements of $\rrnue$  is needed 
to disentangle different solutions at $3 \sigma$ level. 

The quantities $\rrmue$
and $\rrnue$  are (double) ratios of fluxes.  
In real experimental situations one deals with  
the numbers of events of certain type.  
For instance, in the water Cherenkov 
detectors  (the IMB, Kamiokande and Super-Kamiokande) 
one observes $\mu$-like and e-like events as 
the sharp and diffuse single-rings respectively.   
In a calorimetric detector, like SOUDAN,
$\mu$-like and e-like events are identified as 
tracks and showers {\em etc.}.
Moreover,  a number of  
criteria is also implemented 
to select samples of events,  
like suitable cuts on the momenta 
and absence of hits 
in the cosmic muons veto shield.
Therefore, in calculations of numbers of events, 
the fluxes are folded with cross-sections, efficiencies of
detection, misidentification functions {\em etc.}. 
As a result the observed events of different type 
do not correspond to certain underlying reactions. 
In particular, both the charged and neutral current reactions 
can contribute to the same sample of events. 
This makes the analysis of the $NC/CC$ ratio more complicated.  
In next Sections we will identify the samples of events which 
represent the ratio $NC/CC$ closely.

Notice that in principle, 
it should be possible for 
the experimental collaborations to perform
the unfolding of the neutrino fluxes. 
Such a type of analysis was performed
by the Fr\'{e}jus collaboration \cite{Fre95}.

\section{Neutral current induced events}

Let us discuss the possibility to detect the neutral  
current induced events.

The simplest option
would be a study of 
the elastic scattering of the neutrino:
\begin{equation}
\nu_\ell\ N \to \nu_\ell\ N,\ \ \ \ \ \ 
\end{equation}
where $N=n,\, p$ and $\ell=$e, $\mu,\,\tau.$
A recoil proton can  be used 
to track this reaction.  
In water Cherenkov detectors, however, the majority of the recoil 
protons originated from the interactions of  atmospheric neutrinos
is below the threshold for Cherenkov radiation.
Calorimetric detectors do not have  this shortcoming, 
but here the problem  is in  small statistics. 
In fact, only 26 {\em proton events} 
(isolated, highly-ionizing short tracks), with no coincident
hits in the veto shield, had been collected at SOUDAN 2
during 2.83 Kton-years of  exposure 
\cite{SOU97}\ \footnote{The evaluation
of the background is essential, and
requires the study of the depth 
distribution of the events.
If we estimate the background as a $10$\% fraction 
of {\em proton events} with shield hits, 
as for analysis of the {\em track} and {\em shower} 
data samples,  the neutral current signal 
reduces to just 10 events approximatively.}. 
As follows from the simulations \cite{SOU97},   
the hypothesis that the {\em proton events} 
are due to neutral current reaction, 
and there is no oscillation effect is 
in agreement with data.  
However, small statistics 
does not allow one to discriminate 
oscillation scenarios. 

At present, the most promising possibility  
to track the neutral current effects is to study the reaction with  
production of one neutral pion:
\begin{equation}
\nu_\ell\ N \to \nu_\ell\ N\ \pi^0 .
\label{pi0}
\end{equation}
It can be identified as appearance of 
the isolated $\pi^0$ without any accompanying signal
(unless the recoil nucleon is visible).
The decay $\pi^0 \to \gamma \gamma$  gives  
two electromagnetic cascades  which can be  detected 
as two  diffuse rings in the water Cherenkov detectors  
or two showers in the  calorimeter detectors.

The number (the rate of production) of isolated $\pi^0$ 
can be compared {\it e.g.} with 
the number of the e-like (or $\mu$-like) 
events produced by 
charged currents:  
$
\rpie \equiv (\pi^0/\,{\mbox{e-like}})_{data}/
(\pi^0/\,{\mbox{e-like}})_{MC} ,
$
that  minimizes the  theoretical uncertainties
in the neutrino fluxes. 
Here $(\pi^0/\,{\mbox{e-like}})_{MC}$ is the predicted ratio.  
In ideal situation $\rpie$ represents the ratio of the fluxes 
discussed in previous Section, so that 
$\rpie \sim \rrnue.$  
In reality one can not exactly identify a sample of events 
which corresponds to reaction (\ref{pi0}). 
Let us consider this problem for water Cherenkov detectors. 

The reaction (\ref{pi0}) can be detected as the 
``$\pi^0$-{\em event}'' which is determined in the
following way: 

(1) Two isolated electromagnetic cascades should be detected 
as two diffuse rings. 
At high energies the two rings tend to merge, therefore   
an upper cut on the total
momentum permits to optimize the
rings separation. In particular the Super-Kamiokande collaboration 
implements the momentum cut $p_\pi < 400$ MeV. 

(2) The invariant mass of the two identified photons should be 
in certain interval around the mass of the pion,  
say, in the interval  
between $100$ and $200$  MeV. 

To study the oscillation effects one should find separately the  
{\it partial} contributions (in absence of oscillations) 
to the total  rates of events of different type    
from the neutral current 
reactions: $N^\nc$, from the charged current reactions 
induced by $\nue$: $N^\cce$, and  from the charged current reactions 
induced by $\num$: $N^\ccm.$
In presence of oscillations these contributions will 
be modified by the flux suppression factors
determined in (\ref{ncrate}, \ref{ncrate1},  \ref{ncrate2}).

The  reaction  (\ref{pi0}) gives the main  contribution to  
the $\pi^0$-{\em events} sample. However, contributions from other 
reactions are also relevant. 
Let us consider them in order.

The reaction of multi-pion production by the neutral currents:  
$\nubarnu_\ell~ N \to $ $~ \nubarnu_\ell ~  n\, \pi \ N $ 
gives $\pi^0$-{\em event} if only one $\pi^0$ is detected, whereas 
all charged pions are below the Cherenkov threshold, and 
moreover, they 
do not produce secondaries which lead to an observable signal. 
The contribution from these reactions is roughly three times smaller 
than that from (\ref{pi0}). 
Also the neutral current reaction 
$\nubarnu_\ell~ N \to ~ \nubarnu_\ell ~  \pi^{\pm}  N'$ leads to 
$\pi^0$-{\em event} if the $\pi^{\pm}$ undergoes the charge exchange. 
This contribution is, however, small. 
Let us denote by $\ennp^\nc$  the sum of 
all contributions from the $NC$ reactions.

There is a substantial contribution to 
$\pi^0$-{\em event} 
sample from the $CC$ reactions. In particular, 
\begin{equation}
\nubarnu_{\mu}  N \to  \mu^\mp\, \pi^0  N',
\label{mupi0}
\end{equation}
leads to the $\pi^0$-{\em event} if the muon energy is below the 
Cherenkov threshold, and subsequent muon decay does not produce 
an observable signal. Smaller contribution comes from 
reaction $\nubarnu_{\mu}  N \to  \mu^\mp\, \, \pi^{\pm} N $, where 
the pion exchanges the charge in subsequent nuclear interactions, and 
the muon is not detected. Small contribution comes from 
the $CC$ multi-pion reaction. Let us denote 
the sum of all $\num$ $CC$ contributions by 
$\ennp^\ccm .$

There are similar $CC$-reactions induced by $\nue .$
Due to low Cherenkov threshold,  
the probability for the electron to be 
undetected  is smaller and the 
contribution from these reactions to 
$\pi^0$-{\em event} sample,  $\ennp^\cce$,  is also small: $\ennp^\cce 
\ll \ennp^\ccm .$ 

The total rate of the $\pi^0$-{\em events} in absence of oscillations 
equals 
$$
\ennp^0 =  \ennp^\nc + \ennp^\ccm +  \ennp^\cce
$$ 
and the relative 
partial contributions can be estimated  as 
\begin{equation}
\ennp^\nc: \ennp^\ccm :\ennp^\cce
\approx 0.80 : 0.18 : 0.02. 
\label{rel0}
\end{equation}
Notice that this result depends on features of specific experiment 
and event selection criteria. The numbers in (\ref{rel0}) 
have been obtained using the Table 5 
from Kajita review 
\cite{Kajita} and 
correspond roughly to the Kamiokande sub-GeV sample. 
We estimate their uncertainties as $\ennp^\ccm \sim  0.18 \pm 0.05$ 
and $\ennp^\cce \sim 0.02 \pm 0.01 .$ 
Similar results are expected for the Super-Kamiokande. 
In what follows we will use (\ref{rel0}) for illustrative purpose, 
mainly, to estimate the sensitivity of the method. 
Detailed  comparison with experimental data requires 
recalculation of these numbers by 
experimental collaborations.
\vskip0.5truecm

Let us now consider the  influence of different modes of oscillations 
on the rate of the $\pi^0$-{\em events}. The 
$\num \osc \nut$ oscillations  suppress  
$\num CC$ contribution only: 
\begin{equation}
\ennp = \ennp^\nc 
+ \ennp^\cce + P \cdot \ennp^\ccm,   
\label{pitau}
\end{equation}
and the ratio of the rates with and without oscillations can be written as 
\begin{equation}
\ennp/\ennp^0 = 1 - (1 - P) \cdot {\ennp^\ccm}/{\ennp^0}.    
\label{pipi0}
\end{equation}
Since  $\ennp^\ccm$  does not exceed 
20 - 25 \%, one expects the decrease of the 
$\pi^0$-{\em event} 
rate due to oscillations  by 10 \% at most. 

For $\num \osc \nue$ oscillations we get according to 
(\ref{ncrate}, \ref{ncrate1}, \ref{ncrate2})
\begin{equation}
\ennp = \ennp^\nc 
+ (r P - (r - 1)P) \cdot \ennp^\cce +  
(r^{-1} - (r^{-1} -1) P) \cdot \ennp^\ccm.  
\label{pinue}
\end{equation}
So the oscillations suppress the rate of $\pi^0$-{\em events} 
at most by 5 \%. 

In contrast, for the oscillations to the sterile neutrinos 
$\num \osc \nus$ we have 
\begin{equation}
\ennp = \frac{1+r P }{r + 1} \cdot \ennp^\nc 
+ \ennp^\cce + P\cdot \ennp^\ccm.  
\label{pist}
\end{equation}
and the suppression is larger 
than 30 \%.  

To avoid the uncertainties related to 
the absolute values of the neutrino fluxes we will study
the double ratios involving the $\pi^0$-{\em events} and the 
e-like  (or $\mu$-like) events: 
\begin{equation}
\rpie \equiv 
\frac{\ennp/\ennp^0}{\enne/\enne^0}.
\label{doub1}
\end{equation}
The rate of the e-like events without oscillations   
equals
\begin{equation}
\enne^0 = \enne^\cce +  \enne^\nc + \enne^\ccm,   
\label{ne}
\end{equation}
where   $\enne^\cce ,$  $\enne^\nc$, and  $\enne^\ccm$ 
are the partial contributions  
from the  $\nue CC$, $NC$ and  $\num CC$ 
reactions correspondingly. 
The relative contributions 
at the Super-Kamiokande in 
the sub-GeV events sample are: 
\begin{equation}
\enne^\cce : \enne^\nc  : \enne^\ccm \approx 0.90 : 0.08 : 0.02 .   
\label{relat}
\end{equation}
The comparison of different 
Monte Carlo simulations allow us 
to estimate possible 
spread of predictions:  $\enne^\nc = 0.08 \pm 0.03$ and 
$\enne^\ccm = 0.02 \pm 0.01 .$ 

In presence of oscillations the contributions in (\ref{ne}) are modified 
by the flux suppression factors 
(\ref{ncrate}, \ref{ncrate1}, \ref{ncrate2}). In particular, 
for  $\num \osc \nut$ oscillations we get   
\begin{equation}
\enne/\enne^0 = 1 - (1 - P) \cdot \enne^\ccm/\enne^0 . 
\label{ee0}
\end{equation} 
Using  Eqs.\ (\ref{pipi0}) and  (\ref{ee0}) we find $\rpie .$ 
Similarly one can find the double ratios 
for other modes of oscillations. 

In Fig.\ 1 we show the predictions for different modes of oscillations 
in the $\rpie - \rmue$ plot. 
We have used the above formulas with relative contributions  
according  (\ref{rel0}, \ref{relat}). 
Notice that the curves for $\num \osc \nut$  
and $\num \osc \nus$ differ, whereas
the curves for $\num \osc \nus$
and $\num \osc \nue$ oscillations coincide in this plot. 
Indeed, as follows from discussion in Sect.\ 2, 
the value of $\rpie$ changes only if 
there is a $\nut$ component in the 
atmospheric neutrino flux. This statement  is not changed even if 
efficiencies of detection and misidentification of the samples 
are taken into account. Clearly, it will be possible to 
disentangle the case of  $\num - \nut$ oscillations from 
$\num - \nus$ using these ratios, although the difference  
is smaller than in the ideal case of fluxes. For 
$\rmue = 0.6$ we get from the Figure 
$\rpie = 0.93$ and $\rpie = 0.72$ for 
$\num - \nut$ and $\num - \nus$
oscillations correspondingly. 
The ratio of the values of $\rpie$ in 
the two cases is 0.77, to be compared with 
0.73 for $\rrnue$ (\ref{rnue2}).
 
\begin{figure}[t]
\centerline{\epsfig{file=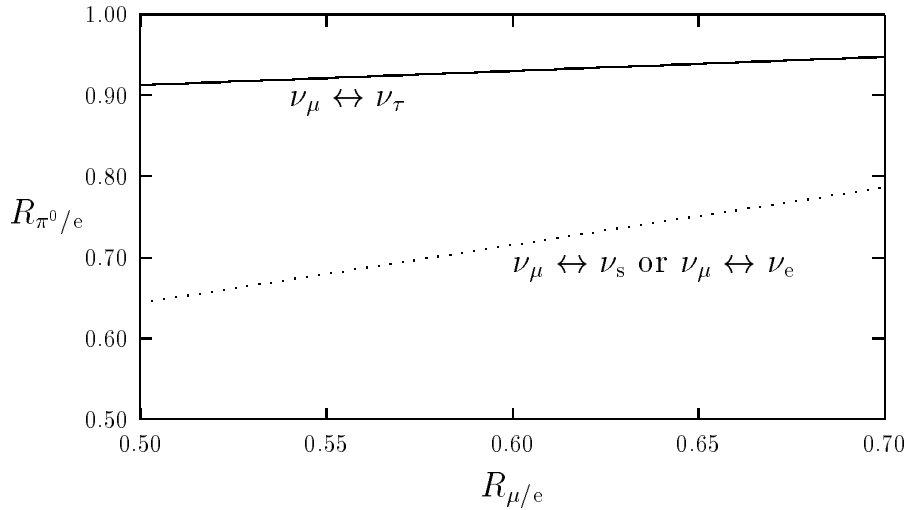,width=12cm}}
\caption{Theoretical dependence on $\rmue$ 
of the $NC/CC$ events ratio $\rpie,$  
under different 
hypotheses of oscillation.}
\end{figure}

Alternatively, one 
can consider the
double ratio $\rpimr$ 
in which the $\pi^0$-{\em events} 
are compared with 
the multi-ring events.
``Inclusive'' data subsets
are numerically rich, 
but their interpretation
is generally less simple.
\vskip0.5truecm

Other samples of data can 
be used in tracking detectors 
with better identification of events  
(SOUDAN or ICARUS).  
Let us consider the samples of two prong events:  
with two showers ($SS$-events), with 
two $\mu$-like tracks ($MM$-events), and  with one 
shower  and one $\mu$-like track  
($SM$-events). These events are generated mainly by 
one pion production reactions.  
$SS$ events are  due to 
$NC$ reaction (\ref{pi0}), 
$SM$ and $MM$ events are due to 
the charged current reactions:
\begin{equation}
\nubarnu_\ell N \to  \ell^\mp\, N\, \pi^\pm,
\ \ \ \ \ \ \ell={\rm e},\,\mu .
\end{equation}
A partial cancellation of the 
uncertainties in the cross-sections 
takes place in the double ratio:
\begin{equation}
\rpism\equiv 
\frac{(\pi^0\,/\,SM)_{data}}{(\pi^0\,/\,SM)_{MC}}
\end{equation}
and in the analogous ratio 
$\rpimm.$ These quantities can be alternatively used 
to estimate the neutral-to-charged-currents ratio. 
Moreover, some uncertainties related to nuclear effects are cancelled in this 
ratios at low energies, where $\Delta$ production mechanism dominates. 
Such a sample can be selected by imposing upper bound on the 
total visible energy. 
A diagnostic of this method is the study of the double ratio 
$\rmmsm=(MM/SM)_{data}/(MM/SM)_{MC},$ 
that in the ideal case should correspond to 
the usual double ratio $\rmue.$

We summarize the relevant reactions, 
the topology of the corresponding events 
and  the rates in the Table 1.  
\begin{table}[bh]
\begin{center}
\begin{tabular}{l l l }   
\hline
\hline
Reaction \ \ \ \ \ \ \ \ \ \  & Type of event \ \ \ \  & Rate  \\ 
\hline
$\nubarnu_{\mu}~N \to\mu^{\mp}~\pi^{\pm}~N$  &$MM$                 & 1.00 \\
$\nubarnu_{\rm e}~N\to {\rm e}^{\mp}~\pi^{\pm}~N$   &$SM$          & 0.43 \\ 
$\nubarnu~N\to~\nubarnu ~\pi^0~N$            &$\pi^0$-{\em events} & 0.27 \\ 
$\nubarnu~N\to~\nubarnu~\pi^+~\pi^-~N$       &$MM$                 & 0.09 \\ 
$\nubarnu_{\mu}~N\to\mu^{\mp}~\pi^{0}~N'$    &$MSY$                & 0.28 \\
$\nubarnu_{\rm e}~N \to {\rm e}^{\mp}~\pi^{0}~N$    &$SSY$         & 0.12 \\ 
$\nubarnu_{\tau}~N\to\tau^{\mp} ~X $         &$M$, $S$ ...         & 0.05 \\ 
\hline 
\hline 
\end{tabular}
\caption[] 
{Reactions of pion production by the atmospheric
neutrinos and their rates normalized to the rate of reaction
$\num~N \to \mu^{\pm}~\pi^{\mp}~N'$,  
$Y = S$ or nothing.} 
\end{center}
\end{table}
The rates were computed for neutrino-nucleon scattering 
using the fluxes and the cross-sections from the 
reports \cite{MC} and \cite{MC1}. 
No efficiencies of detection have been taken into account. 

According to the Table 1 there are two contributions to 
$SS$ events (in assumption of good efficiency of detection): 
from $\pi^{0}$ production by the $NC$ (reaction 3) and from 
$\nue CC$ reaction with production of the $\pi^{0}$ (reaction 6),  
when $\pi^{0}$ is detected as one showering event. 
The latter reaction has more than two times smaller rate and its 
contamination can be further suppressed  by  imposing 
the  invariant mass criteria and 
the upper cut on  visible energy. 
The $SM$ sample also has two contributions: from $\nue CC$ production 
of the charged pion (reaction 2) and from $\num CC$ production of the 
$\pi^{0}$ (reaction 5), when $\pi^{0}$ is detected as one showering event.
Again the latter contribution can be suppressed by 
an upper cut of the visible energy. Using the numbers from the Table 1 we 
conclude that    
the double ratio $\rpism$ indeed can represent 
the ratio of $NC/CC$ provided the products 
of reactions are well identified.  

However in tracking detectors with 
$^{40}$Ar (ICARUS) or 
iron (SOUDAN) as targets, 
nuclear effects should be taken into account. 
Charge exchange leads to further mixture of samples, moreover, 
if production of $\Delta$ occurs in heavy nuclei,  
the emission of pions is mainly a surface phenomenon. 
Certainly, more studies are needed 
to show the validity of the method for complex nuclei target 
with nonzero isospin. 

\section{Neutral Pions at Super-Kamiokande}

Let us perform a tentative analysis of $NC/CC$ ratios 
using available experimental data. 
The Super-Kamiokande has already collected 
a good statistics of {\em single $\pi^0$ events},
defined as the events with 
(1) only two rings, both of electromagnetic type
($SS$-events),
(2) reconstructed vertex in the fiducial volume,
(3) total momentum smaller than 400 MeV
\cite{Kas98}.
The study of the distributions in the 
invariant mass $m_{\pi^0}$ 
was performed in order to calibrate
the energy measurements at the  
Super-Kamiokande \cite{SK}. 

Additional criteria could be implemented to diminish 
the contribution of the $CC$ reaction (\ref{mupi0}) 
to the  $\pi^0$-{\em event} sample. 
This reaction gives a $\pi^0$-{\em event} if the muon 
is below the Cherenkov threshold. 
This muon can still be observed if it decays in the detector. 
Thus one can require an absence of the $\mu$-decay in appropriate 
time window.

In what  follows, we will use the 
results from  the 20 Kton-year exposure and 
from the Monte Carlo  simulations which 
correspond to 224.6 Kton-year 
\cite{Kas98}. 
Let us  compare the observed number of neutral pion 
with the expectations.
We will use narrower  
window of the invariant mass of two photons:   
$m_{\pi^0}=100-200$ MeV.  
{}From Fig.\ A3 in \cite{Kas98} we get  
in this window 
the numbers:    
$\ennp^{exp} = 72$  observed $\pi^0$-{\em events}  events 
and $\ennp^{MC} = 55$ expected events (the Monte Carlo data 
have been normalized to the same exposure time as observations).
Extrapolating the distribution of events from the regions outside the peak to 
the peak region we can subtract the background. 
The above  numbers of events become: $\ennp^{exp} = 58$ and 
$\ennp^{MC} = 41.6 .$ 
That is, the number of observed $\pi^0$-{\em events} 
exceeds the expected number. 

In the Table 2 we present  the data-to-Monte Carlo 
ratios for the  events of different  type 
with and without oscillations. 
The MC predictions in presence of oscillations 
have been calculated according to formulas 
(\ref{pitau}, \ref{pinue}, \ref{pist}, \ref{ee0}).  
For $\pi^0$-{\em events} we used relative 
elementary contributions from
$\num CC,$ $\nue CC$ and $NC$ reactions 
from Eq.\ (\ref{rel0}), and  
for e-like, $\mu$-like and multi-ring
events we have taken the elementary contributions from  
\cite{Kas98bis}. The flux suppression factors were evaluated for 
$P=0.6$ which permits to account for 
the double ratio $\rmue .$  

\begin{table}[t]
\begin{center}
\begin{tabular}{l c c c}   
\hline
\hline
data/MC  &           no oscill. & $\num\osc\nut$ & $\num\osc\nus$ \\
\hline
e-like                                     & 1.21 & 1.22   & 1.24 \\  
$\mu$-like                                 & 0.77 & 1.25   & 1.27 \\
multi-ring                                 & 0.89 & 1.07   & 1.20 \\
$\pi^0$-{\em events}                       & 1.31 & 1.42   & 1.86 \\ 
$\pi^0$-{\em events}, bkgr.\ subtr.\ \ \ \ & 1.39 & 1.51   & 1.98 \\
\hline
\hline
\end{tabular}
\caption[] {Data-to-Monte Carlo event ratios
under different hypotheses of oscillation.}  
\end{center}
\end{table}

According to Table 2 
in absence of oscillations
there is a shortage of $\mu$-like and an excess
of e-like events; the latter excess however 
could be explained by
theoretical uncertainties in the 
normalization of the flux.
The number of observed $\pi^0$-{\em events} 
exceeds the numbers expected 
under any hypothesis, moreover the oscillations, 
especially $\num \osc \nus$ 
even enhance the difference. 

The double ratios are presented in Table 3.
\begin{table}[b]
\begin{center}
\begin{tabular}{l l l l}   
\hline
\hline
         &  no oscill.  & $\num\osc\nut$ & $\num\osc\nus$ \\
\hline
$\rpie$  &  1.08 (1.15) & 1.16 (1.24)    & 1.50 (1.60)    \\
$\rpimr$ &  1.47 (1.56) & 1.33 (1.41)    & 1.55 (1.56)    \\
$\rmue$  &  0.64        & 1.02           & 1.02           \\
\hline
\hline
\end{tabular}
\caption[] {Double ratios
under different 
hypotheses of oscillation. 
In bracket we present the values after the 
subtraction of the background for  the 
$\pi^0$-{\em events}.}  
\end{center}
\end{table}
As follows from the Table, the best agreement with the data 
is obtained under the $\num \osc \nut$ 
oscillation hypothesis 
(with a 15-25\% excess of $\pi^0$-{\em events}). 
Let us estimate  uncertainties in this analysis.  
The  experimental (statistical) errors is  
related  mostly to the small 
number of observed $\pi^0$-{\em events}. 
With present data it is about  15\%, and it will decrease 
with  the accumulation of statistics.
The theoretical uncertainties are much larger. 
The results of two calculations of the neutrino-induced 
single-$\pi$ production cross-sections 
by Fogli-Nardulli \cite{FN} 
and Rein-Sehgal \cite{RS} 
differ by  20\% \cite{MC}.
Using the results in 
Sec.\ 3.2 of \cite{MC} we estimate 
the  uncertainties related to the nuclear effects
as being about the 20\% .
So, the overall uncertainty could be  
around 35-40\%. 
Already this uncertainty is larger than the differences in 
Eq.\ (\ref{range}) or in Fig.\ 1, 
which means  that at present it is impossible 
to exclude different channels of oscillations. 

On top of this 
there is   a possible 
neutron-induced background \cite{Rya94} 
which can contribute significantly  to  the $\pi^0$ data.
This  background is more important 
for  $\pi^0$ sample than for e-like sample considered in \cite{Kam96} 
for two reasons: 
(1) smaller statistics of the $\pi^0$  events; 
(2) the absence of suppression  which exists 
for  the e-like events, since only in  17\%
of the cases the  neutron-produced 
neutral pions can fake an electron \cite{Kam96}. 
The neutron background 
at Kamiokande was below 30\% .
Due to large size of the detector and the possibility to use central 
parts of the detector, the Super-Kamiokande can reduce 
possible effect of neutrons up to desired level.

\section{Conclusions and Perspectives}

The study of neutral current observables, 
and in particular of the 
$\pi^0$-{\em events}, can provide  deeper 
insights in  to the atmospheric neutrino problem. 
This study will give not only new independent check of the oscillation 
hypothesis, but also will allow one to discriminate between 
different channels of oscillations.

An exciting perspective is related to the 
Super-Kamiokande operations. Already to the end of 1998
it will collect more that 300 
$\pi^0$-{\em events} \cite{Suz96}. 
Our tentative analysis of the recent SK data on  
neutral pions indicates  a slight preference of  
the hypothesis of $\num\osc\nut$ 
oscillations, even though (as was emphasized) 
the  uncertainties are  large.

Further data taking at  the Super-Kamiokande will 
lead to desired decrease of the experimental errors. 
As far as  predictions are concerned,  
the progress could come 
not only from new updated 
series of calculations 
\cite{singh} but also from the future 
experimental studies of the neutrino induced 
single pion emission reactions at GeV energies.
These studies will be possible 
with  the two front detectors of 
the K2K experiment \cite{K2K}, 
that will start to operate in 
the beginning of 1999 \cite{K2Kbis}.
The K2K collaboration plans 
also to search for the oscillation 
effects using $\pi^0$-{\em events} directly 
\cite{K2K}. The ratio of events ($\pi^0/\mu$-like)
in the front detector and in the Super-Kamiokande 
detector will be measured. 
However, it is not guaranteed 
that oscillation effects will be observable
with the range of values of 
$\Delta m^2$ preferable 
now \cite{SK,K2Kbis}.
\vskip0.3cm
\noindent{\Large\bf Acknowledgments} 
\vskip 0.3cm
The authors are 
grateful to S.P.\ Mikheyev for 
illuminating discussions. 
F.\ V.\ acknowledges
with pleasure conversations with 
F.\ Cafagna, K.\ Daum, M.\ Goodman,
E.A.\ Paschos and B.\ Saitta; 
he thanks Balaji and 
S.\ Uma Sankar for suggesting 
to consider the multi-flavour case. 
A.\ S.\ is grateful for 
discussions to S.K.\ Singh and 
M.J.\ Vicente-Vacas. 

The first version of this paper has been written during 
the extended workshop ``Highlights in Astroparticle Physics'',
ICTP, Trieste, October 15 - December 15, 1997.


\begin{thebibliography}{99}

\bibitem{SK} Y. Fukuda et al., 
(Super-Kamiokande collaboration) 
ICRR-411-98-7, hep-ex/9803006. 

\bibitem{SOUDAN} T. Kafka (for the SOUDAN 2 collaboration)
talk at TAUP 97, September 1997, LNGS, Assergi, Italy,
hep-ph/9712281.

\bibitem{prob1} K.S. Hirata et al.
(Kamiokande collaboration), Phys.Lett. B205 (1988) 416; 
ibidem, B 280 (1992) 146.  
\bibitem{prob2} D. Casper et al. (IMB 
collaboration), Phys.Rev.Lett. (1991) 2561;  
R. Becker-Szendy et al., Phys.Rev. D46 (1992) 3720. 

\bibitem{atm1} J.G. Learned, S. Pakvasa 
and T.J. Weiler, Phys.Lett. B207 (1988) 79.
\bibitem{atm2} V. Barger and K. Whisnant, 
Phys.Lett. B209 (1988) 365.
\bibitem{atm3} K. Hidaka, M. Honda and 
S. Midorikawa, Phys.Rev.Lett. 61 (1988) 1537.

\bibitem{reac1} B. Achkar et al.,
Nucl.Phys. B434 (1995) 503.
\bibitem{reac2} G.S. Vidyakin et. al., 
JETP Lett. 59 (1994) 364. 
\bibitem{reac3} M. Apollonio et al., 
(CHOOZ collaboration), hep-ex/9711002 

\bibitem{steril} E. Akhmedov, 
P. Lipari and M. Lusignoli, 
Phys.Lett. B300 (1993) 128;
E. Akhmedov, Proc. of the Int. School
on Cosm. Dark Matter,  
eds. J.W.F. Valle and A. P\'erez (Valencia 1993).

\bibitem{steril2} R. Foot, R.R. Volkas and O. Yasuda
hep-ph/9801431. 

\bibitem{smi} A.Yu. Smirnov, 
Proc. of the 28$^{th}$ Int. Conf. 
on High Energy Physics, eds.  
Z. Ajduk and A.K. Wroblewski, 
p. 288 (1997); hep-ph/9611465. 
\bibitem{liu} Q.Y. Liu and A.Yu. Smirnov,
hep-ph/9712493, 
to appear in Nucl.Phys. B.
For a detailed comparison
with the experimental data 
see \cite{fluxint}.

\bibitem{fluxint} Q.Y. Liu, S.P. Mikheyev and
A.Yu. Smirnov hep-ph/9803415; 
P. Lipari and M. Lusignoli, hep-ph/9803440.

\bibitem{combine} O.  Yasuda,  TMUP-HEL-9603,
hep-ph/9602342,  TMUP-HEL-9706, hep-ph/9706546;  
Mohan Narayan, G. Rajasekaran and S. Uma Sankar, 
Phys.Rev. D56 (1997) 437;  
G.L. Fogli, E. Lisi, D. Montanino and 
G. Scioscia, Phys.Rev. D55 (1997) 4385.  

\bibitem{pakv} J.W. Flanagan, J.G. Learned and S. Pakvasa, 
Phys.Rev. D57 (1998) 2649; 
R. Foot, R.R. Volkas and O. Yasuda,  hep-ph/9710403. 

\bibitem{Fre95} K. Daum et al.,
(Fr\'{e}jus collaboration), 
Z.Phys. C 66 (1995) 417.

\bibitem{SOU97} H. Gallagher
for the SOUDAN 2 collaboration,
talk given at the
XVI International
Workshop on Weak Interactions and 
Neutrinos (WIN97) Capri 1997. 

\bibitem{Kajita} T. Kajita, contribution
to the book ``Physics and astrophysics of neutrinos'', 
M. Fukugita and A. Suzuki eds., 1994.

\bibitem{MC} M. Nakahata et al.,
Journ.Phys.Soc. Japan 55 (1986) 3786. 

\bibitem{MC1} M. Takita , ICR-Report-186-89-3. 

\bibitem{Kas98} S. Kasuga, 
University of Tokyo Ph.D.\ thesis,
January 1998.

\bibitem{Kas98bis} See Table 7.1
in reference \cite{Kas98}. The results
of the simulations are fairly consistent 
with those in \cite{SK},
but slightly different from those of the 
Kamiokande Monte Carlo, especially 
for the $NC$ induced e-like signal
(see Table 5 of \cite{Kajita}).

\bibitem{FN} G.L. Fogli and G. Nardulli, 
Nucl.Phys. B160 (1979) 116; 
Nucl.Phys. B165 (1979) 162. 
\bibitem{RS} D. Rein and L.M. Sehgal, 
Ann.Phys. 133 (1981) 79.


\bibitem{Rya94} O.G. Ryazhskaya, 
JETP Lett. 60 (1994) 617;
ibidem 61 (1995) 237.

\bibitem{Kam96} Y. Fukuda et al.,  
(Kamiokande collaboration), 
Phys. Lett. B 388 (1996) 397.
In this work, the observed 
neutrino-induced neutral-pion 
signal was used to study the 
background to the e-like event 
sample which could be 
related to neutrons \cite{Rya94}.
The expected number of $\pi^0$ 
events was not presented there, 
since the Monte Carlo results 
were normalized to the data.

\bibitem{singh} S.K. Singh, M.J. Vicente-Vacas, E. Oset, 
Phys.Lett. B 416 (1998) 23. 

\bibitem{Suz96} Y. Suzuki, in the Proc. of the 
17$^{th}$ Int. Conf. on Neutrino Ph. and Astroph. 
(Neutrino '96), eds. K. Enqvist, 
K. Huitu and J. Maalampi,
World Scientific, 1997. 
However, the definition of 
$\pi^0$-{\em events} used 
is not indicated.

\bibitem{K2K} After releasing a 
first version of the present paper, 
we became aware that the idea 
to track the neutral current effects 
by using the neutral pions 
was first suggested in the preprint:
K. Nishikawa, INS-Rep.-924 
(1992) and then in \cite{Suz96}.

\bibitem{K2Kbis} See the recent talk of 
Y. Oyama (for the K2K collaboration),
hep-ex/9803014.
\end{thebibliography}
\end{document}